\documentclass[prd,superscriptaddress,nofootinbib, showkeys, 
preprint]{revtex4}
\usepackage{graphicx}
\usepackage[usenames,dvipsnames]{color}
\usepackage{amsmath,amssymb}
\usepackage{mathtools}
\usepackage[colorlinks]{hyperref}
\usepackage{float}

\newcommand{\be}{\begin{eqnarray}}                                             
\newcommand{\ee}{\end{eqnarray}}
\newcommand{\nn}{\nonumber}

\newcommand{\noplus}{}

\newcommand{\tmop}[1]{\ensuremath{\operatorname{#1}}}
\begin{document}
	\title{   Deconfinement to confinement by generalizing BRST symmetry on the sphere}
		
	\author{Haresh Raval}
	\email{haresha.raval@juitsolan.in} 
	\affiliation{Department of Physics and Materials Science, Jaypee University  of Information Technology,
	Solan 173234, Himachal Pradesh, India}
	
	\begin{abstract}
 Recently it has been shown that the theory  in the quadratic gauge on  4-sphere, $\mathbb{S}^{4}$ consists of
  two phases namely, the  confined  and the  deconfined phases. 	A suitable  finite field dependent BRST (FFBRST)  transformation  interrelates  two different gauge fixed theories. In this paper, we use the FFBRST technique on the curved space for the first time and  elaborate a  novel application of it.  We 
   propose  two different formulations of this technique that transform the deconfined phase action on sphere to  the confined phase action on sphere inside the quadratic gauge. Both proposed passages  change the phase with BRST  invariance  to the phase without BRST  invariance unlike usual connections  
   where  the   FFBRST operation leave the BRST symmetry intact and there is  a unique
   field theoretic essence of them, which makes them particularly important to study.  Thus, the two different  field redefinitions act as a new mechanism that execute phase transition between two real QCD phases on 4-sphere other than ghost condensation process.

  	\end{abstract}
  \maketitle
	\section{Introduction}
	Consequences of  compactness of the manifold on the phenomena in  gauge theories are relevant physically as indicated by  experimental evidences (Ref.~\cite{@} and refs. in that). Therefore, the research in gauge theory  on the sphere has generated a lot of interest as the sphere is a simple compact manifold and a theory  on the sphere lends a model to study effects of compactness of  the manifold on phenomena in QCD.
In ref. \cite{1} which  formulates the massless quantum electrodynamics on a sphere in Euclidean space with 5 dimensions, it was asserted that compactification
of the space makes  this theory infra-red finite. Thereafter, gauge theories on a hypersphere with manifest $O(n)$-covariance
have been investigated  in different contexts~\cite{2,3,4,5,5a,5a',5b}. 
 For example, the application of conformal Killing vectors in  constructing the
$O(n)$-covariant formulation was demonstrated in Ref.~\cite{5a'}.  

	 
	 The quadratic gauge which is found to have deep non perturbative implications is given as follows in the Minkowski space~\cite{6}
	 	\begin{align} \label{eq:0}
	 A^a_{\mu} ( x) A^{\mu a} ( x) = f^a ( x) ; \  \text{  for each $a$, }
	 \end{align}
	 where $f^a(x)$ is an arbitrary function of $x$. The corresponding Faddeev-Popov action is given as
	 \begin{eqnarray} \label{eq:Leff}
	 \mathcal{L}_{Q}	 = - \frac{1}{4} F^a_{\mu \nu} F^{\mu \nu a}
	 \noplus - \frac{1}{2 \zeta} \sum_a  ( A^a_{\mu} A^{\mu a})^2 -2\sum_a \overline{c^a}
	 A^{\mu a} ( D_{\mu} c)^a  ,
	 \end{eqnarray}
	 where $c, \bar{c}  $ are ghost and anti-ghost fields respectively,   $\zeta$ is an arbitrary gauge  parameter,  the field strength $F^a_{\mu \nu}= \partial_{\mu}A^a_{\nu}(x)- \partial_{\nu}A^a_{\mu}(x)-g 
	 f^{abc} A^b_{\mu}(x)A^c_{\nu}(x)$   and $(D_{\mu} c)^a = \partial_\mu c^a - g f^{a b c} A_\mu^b c^c$.   The  indices $a$, $b$ and $c$ are  independently summed over   $1$ to $N^2-1$ in Eq.~\eqref{eq:Leff}. It
	  has been rigorously studied in various frameworks in the recent past \cite{6,7,8,9,10,11,epl,101,111}. To mention a few of them,
	  we showed that the quadratic gauge fixed  theory can be transformed into the effective theory in Lorenz gauge under
	  proper field redefinitions~\cite{8}.  In ref.~\cite{9}, we applied the quadratic gauge to SO(N) QCD  to probe the infra-red behavior of the theory.  We constructed two  superspace versions of this same theory, one employs a new and extended BRST symmetry~\cite{10}, and the other was developed without full anti-BRST symmetry~\cite{epl}.

	  
	    The role of BRST transformation  in  quantizing the gauge theories is crucial. 
	    The infinitesimal  anti-commuting  parameter of a usual BRST transformation can be generalized to be finite and field dependent  leaving the form of a transformation unchanged as was first
	  done   in ref.~\cite{12}.  The explicit dependence on space time coordinates is absent in  the generalized parameter.    The field dependence of such FFBRST transformations varies the path integral
	  measure keeping
	  other characteristics  of 
	  infinitesimal BRST transformations intact. Therefore,  the generating
	  functional of a BRST invariant theory loses invariance
	  under FFBRST. The change in the measure introduces a Jacobian in the path integral which depends on generalized parameter. Thus, the action that represents the Jacobian acts as a factor which can convert one effective theory to other. Therefore, FFBRST  may 
	  be useful to get an insight into  phenomena in one theory given the knowledge of the same in the other theory. Hence, this technique finds  numerous applications in interrelating   two BRST invariant gauge fixed theories~\cite{13,14,15,16,17,18}.

 In this paper,   two novel FFBRSTs are suggested that
   transform the    action in the deconfined  phase on 4-sphere to that in the confined phase on 4-sphere. The suggested passages  carry  a unique
physical significance that they execute transition between two real QCD phases in compact space and convert phase with BRST invariance to phase without BRST  invariance unlike the usual cases. Before we proceed to the proposal in Sec. IV,  Yang-Mills theory on a general sphere is reviewed in the next section.     In Sec. III however only 4-sphere, $\mathbb{S}^{4}$  is considered so that the study is rendered physically relevant and  the  phenomenon of  ghost condensation in the current theory on 4-sphere is reexamined.

\section{ QCD on hypersphere}
Here we revise the  structure of QCD on  hypersphere, $\mathbb{S}^{n-1}$ of $ (n-1)$ dimensions with manifest $O(n)$ covariance.   The Euclidean space, $\mathbb{R}^{n}$ embeds this sphere.    The sphere $\mathbb{S}^{n-1}$  is considered to have a unit radius  i.e.,  $r_\alpha r_\alpha =1$, where $r_\alpha$ is a cartesian coordinate of a point on the sphere $\mathbb{S}^{n-1}$ and $\alpha = 1,2,...,n$. 
  The spherical symmetry of the underlying space implies that the angular momentum  operator governs dynamics of a theory on sphere, which is given by
\be\label{de92}
       L_{\alpha \beta}   = -i \big( r_\alpha \partial_\beta-r_\beta \partial_\alpha\big);\ \    \partial_\beta \equiv \frac{\partial}{\partial r_\beta},\ \  \beta, \alpha = 1,2,...,n.
       \ee
Particularly,
\be\label{de91}
L_{\mu n} = -L_{n \mu}=  i r_n \partial_\mu,
\ee
as $ \frac{\partial}{\partial r_n}$ is zero by definition regardless of  what it operates on since the variable $r_n$ is not independent.
The operator $ L_{\alpha \beta}$ obeys the following Lie algebra
\be\label{al}
[L_{\alpha \beta},L_{\gamma\eta} ] = i(\delta_{\alpha\gamma}L_{\beta\eta} -\delta_{\beta\gamma}L_{\alpha\eta} - \delta_{\alpha\eta}L_{\beta\gamma} +\delta_{\beta\eta}L_{\alpha\gamma}).
\ee
The   stereographic coordinates  on the hyperplane  $\mathbb{R}^{n-1}$ that bisect the sphere are $\{y_\mu, \mu = 1,2,...,n-1\} $. These stereographic coordinates map coordinates of the point on  sphere, $r_\mu$ as 
\be\label{st1} 
r_\mu = \frac{2 y_\mu}{1+y^2},\ \ \  r_n = \frac{1-y^2}{1+y^2},\ \ \  y^2 \equiv y_\mu y_\mu.
\ee
 The gauge field $\bar{A}$ on the sphere, $\mathbb{S}^{n-1}$  and   the  field $A$ on the stereographic plane are related as follows \cite{5a'}
\be\label{st2}
\bar{A}_\mu (r) = \frac{1+y^2}{2} A_\mu(y) -y_\mu y_\nu A_\nu(y), \ \ \bar{A}_n  =-y_\mu A_\mu;\ \ \mu, \nu = 1,2,...,n-1.
\ee
The overbar appearing through out this paper indicates fields on sphere. In Euclidean frame,  upper and lower indices signify the same tensor.    The gauge field $\bar{A}$ is Lie algebra valued i.e., $\bar{A}_\mu= \bar{A}_\mu^a T^a$, where\ $ T^a$ is a generator of  SU(N) group.   The stereographic projections in Eqs.~\eqref{st1}, \eqref{st2} leads to the following  transversality condition  on sphere
\be
r_\alpha \bar{A}_\alpha =0 \Rightarrow r_\alpha \bar{A}_\alpha^a=0,\ \  \alpha = 1,2,...,n.
\ee Therefore, this condition is   inherent to the QCD on hypersphere.   This implies that  the gluon $\bar{A}^a$  are tangential to $\mathbb{S}^{n-1}$.  

The form of a gauge transformation relies on the underlying spatial geometry in which
theory is laid down.
The gauge transformation  on $\mathbb{S}^{n-1}$ is  the following
\be \label{3}
\delta \bar{A}^a_\beta =  r_\alpha(r_\alpha D_\beta-r_\beta D_\alpha) \epsilon^a =  r_\alpha i \mathcal{L}_{\alpha \beta }\ \epsilon^a,  \ee
where $\epsilon$ is a parameter of the transformation, $D_\beta \epsilon^a =  \partial_\beta \epsilon^a -g f^{abc} \bar{A}_\beta^b \epsilon^c$. We see that  the  operator $ \mathcal{L}_{\alpha \beta }$ is a covariantized  rendition of angular momentum   owing to the local symmetry. Now, $r_\beta \delta \bar{A}^a_\beta= 0$  which affirms that an infinitesimal gauge transformation is also tangent to  $\mathbb{S}^{n-1}$.    Eq.~\eqref{3} can further be simplified as below
\be \label{4}
\delta \bar{A}^a_\beta =   ( D_\beta- r_\alpha r_\beta D_\alpha) \epsilon^a, \ \ \text{ because } \ r_\alpha r_\alpha =1.
\ee
 The gauge invariant Yang-Mills action on $\mathbb{S}^{n-1}$  under the transformation in Eq. \eqref{4}  can be given by
\be \label{8}
S_{YM} = -\frac{1}{12} \int d \Omega  \ \bar{F}_{\alpha \beta \gamma }^a \bar{F}_{\alpha \beta \gamma }^a,
\ee
where 
\be
\bar{F}_{\alpha \beta \gamma }^a = r_\gamma   \bar{F}_{\alpha\beta} ^a + r_ \beta   \bar{F}_{ \gamma \alpha } ^a   +  r_ \alpha   \bar{F}_{ \beta \gamma} ^a,
\ee  which is  the $O(n)$ covariant rank-3 tensor and 
\be 
d \Omega= \frac{1}{|r_n|} \displaystyle\prod\limits_{\mu=1}^{n-1} d r_\mu
\ee
is an invariant measure on $\mathbb{S}^{n-1}$.
 
	\section{ Ghost condensation as  a mechanism for confinement  on 4-sphere }
 	 The deconfinement to confinement phase transition on 4-sphere due to the  ghost condensation  is a crucial study
 for the present context which we now review. The generalization to higher dimensions is obvious.     We introduced the  quadratic gauge and the related Faddeev-Popov operator $ \Delta_{FP}$ on $\mathbb{S}^{4}$  respectively as follows \cite{111}    
	\be 
	&&\bar{A}^a_\beta (r)  \bar{A}^a_\beta (r) = f^a(r); \ \ \beta = 1,2,...,5, \ \text{for each $a$,}\\
 \Delta_{FP}&=&  \det\left[2\bar{A}^a_\beta\Big( \partial_\beta \delta^{ab}-g f^{acb}\bar{A}_\beta^ c- r_\alpha r_\beta(\partial_\alpha \delta^{ab}-g f^{acb}\bar{A}_\alpha^ c)\Big)\right] .
	\ee  
	Therefore,  the gauge fixing and the ghost actions turn out to be as shown below
	\be\label{10}
	S_{GFQ} +S_{ghostQ} =  \int d \Omega \left[ -\frac{1}{2\xi} (\bar{A}^a_\beta  \bar{A}_\beta^a)^2-2 \hat{\bar{c}}^a \bar{A}_\beta^a D_\beta \bar{c}^a+2 \hat{\bar{c}}^a \bar{A}_\beta^a  r_\alpha r_\beta D_\alpha \bar{c}^a  \right],
	\ee
		$d\Omega$ is the  4-sphere angular measure.   Now, the built-in  transversality eliminates the last
		 term in Eq.~\eqref{10}. Therefore, we have
		\be\label{28}
		S_{GFQ} +S_{ghostQ} =  \int d \Omega \left[  -\frac{1}{2\xi} (\bar{A}^a_\beta  \bar{A}_\beta^a)^2-2 \hat{\bar{c}}^a \bar{A}_\beta^a D_\beta \bar{c}^a\right],\ \ \ \text{ since  } \  r_\beta \bar{A}_\beta^a=0.
		\ee
		 The resulting  quadratic gauge fixed action on $\mathbb{S}^4$  is the following \be S_{eff}= \int d \Omega \left[ -\frac{1}{12} \ \bar{F}_{\alpha \beta \gamma }^a \bar{F}_{\alpha \beta \gamma }^a -\frac{1}{2\xi} (\bar{A}^a_\beta  \bar{A}_\beta^a)^2-2 \hat{\bar{c}}^a \bar{A}_\beta^a D_\beta \bar{c}^a\right].  \ee
		 This action  holds no indication of confinement, it signifies perturbative regime just like any other usual action. Therefore, it characterizes the normal or deconfined   phase on $\mathbb{S}^{4}$.
 
 	Let us rewrite the ghost Lagrangian as follows
	\be\label{gh}
	2 \hat{\bar{c}}^a \bar{A}^{ a}_\beta  D_\beta \bar{c}^a = 2 \hat{\bar{c}}^a \bar{A}^{ a}_\beta \partial_\beta \bar{c}^a -2	gf^{abc}   \hat{\bar{c}}^a {\bar{c}}^c \bar{A}_\beta^b \bar{A}_\beta^a. 
	\ee
	In the  ghost condensed state, the vacuum expectation value of the first term on right hand side vanishes \cite{111}.  
The second  term  gives the mass matrix for gluons on $\mathbb{S}^{4}$  as follows
 	\begin{equation}
 (M^{ 2})^{a b}_{\tmop{dyn}} = 2 g \displaystyle\sum\limits_{c=1}^{N^2-1}f^{a b c} 
 \langle\hat{\bar{c}}^a \bar{c}^c\rangle.
 \end{equation}
 In the   state with all ghost-anti{\small -}ghost 
 condensates   identical i.e., 
 \begin{equation}\label{13}
 \langle\hat{\bar{c}}^1 \bar{c}^1\rangle = ... =  \langle\hat{\bar{c}}^1 \bar{c}^{N^2-1}\rangle = ... =  \langle\hat{\bar{c}}^{N^2-1} \bar{c}^1\rangle = ... =     
 \langle\hat{\bar{c}}^{N^2-1} \bar{c}^{N^2-1}\rangle = K',
 \end{equation}
 an interesting case emerges. 	Before proceeding to that, it is useful to comment on BRST variation of the condensate. Variation of the condensate under BRST is given as $\delta\langle (\hat{\Bar{c}}^m \Bar{c}^n) \rangle =\langle \delta(\hat{\Bar{c}}^m \Bar{c}^n) \rangle =\omega\Big( \langle \Bar{B}^m \Bar{c}^n\rangle +\frac{1}{2}f^{nij} \langle \hat{\Bar{c}}^m \Bar{c}^i\Bar{c}^j \rangle \Big)$ (as $\delta\hat{\Bar{c}}^m  = \omega\Bar{B}^m $ in terms of auxiliary field $\Bar{B}$).  Since Coleman Weinberg mechanism does not give rise to terms with ghost number 1 nor the present theory contains such terms, condensates $\langle \Bar{B}^m \Bar{c}^n\rangle$ and $ \langle \hat{\Bar{c}}^m\Bar{c}^i\Bar{c}^j \rangle $ must be identically zero hence $\delta\langle (\hat{\Bar{c}}^m \Bar{c}^n) \rangle =0$ which is consistent with Eq.~\eqref{13}.
 
 The mass matrix in this state now simplifies to the following
 \begin{equation}\label{mm}
 (M^{ 2})^{a b}_{\tmop{dyn}} = 2 g \displaystyle\sum\limits_{c=1}^{N^2-1}f^{a b c} K'.
 \end{equation}
This mass matrix  is anti-symmetric and has $N(N - 1)$ non-zero eigenvalues, which  implies that $N(N - 1)$ off-diagonal gluons on sphere obtain mass  $M_{gluon} = \frac{1}{\sqrt{2}}(1\pm i)m$ and the rest $N-1$ diagonal gluons on 4-sphere remain massless.     So in this description, 
 only the diagonal gluons mediate interactions in the Infra-red region which are long ranged. This firmly indicates presence of  Abelian dominance on the sphere.  We demonstrated in Ref. \cite{6,9} that Abelian dominance exists in   Euclidean space too in the  quadratic gauge fixed theory.   Thus, infrared sector of the quadratic gauge in $\mathbb{R}^4$  and  on $\mathbb{S}^4$  is  alike.   
 
 		Realizing the identity in Eq.~\eqref{13} within a legitimate mechanism on the 4-sphere is important to retain physical relevance of the infra-red consequence of the theory.  This can be done in the following manner.
  Let's  first see the following map between ghost fields on sphere and corresponding ghosts in the  Euclidean space~\cite{5a}
 \be\label{mmap} 
 \bar{c}^a (r) = \frac{1 + y^2}{2} {c^a}(y), \ \ \hat{\bar{c}}^a (r) = \frac{1 + y^2}{2} \tilde{c}^a(y),
 \ee
 here $ c(y), \tilde{c}(y)$ are ghost and anti-ghost fields respectively in the flat space. We put this map in
 following identity for ghost condensations in the flat spacetime $\mathbb{R}^4$  whose demonstration  within  Coleman Weinberg mechanism is  provided  in \cite{6,9},
 \be\label{id}
  \langle{\tilde{c}}^1 {c}^1\rangle = ... =  \langle{\tilde{c}}^1 {c}^{N^2-1}\rangle = ... =  \langle{\tilde{c}}^{N^2-1} {c}^1\rangle = ... =     
 \langle{\tilde{c}}^{N^2-1} {c}^{N^2-1}\rangle = K (Const.).
 \ee
  As a result, 
 	the matrix in Eq.~\eqref{mm} becomes 
 \be\label{mm'}		(M^{ 2})^{a b}_{\tmop{dyn}} = 2 g \displaystyle\sum\limits_{c=1}^{N^2-1}f^{a b c} K'=  2  g  \big[ \frac{1 + y^2}{2}\big]^2  \Big(\displaystyle\sum\limits_{c=1}^{N^2-1} f^{a b c} K \Big).
 \ee
  We need to diagonalize Eq.~\eqref{mm'} to find  the mass squared of a gluon on $\mathbb{S}^4$,  $M_{a}^2 $ which is as following 
 \be \label{mr}
 M_{a}^2 =2 g  \big[ \frac{1 + y^2}{2}\big]^2  m_{a}^{*2} = (1 +r_5)^{-2} m_a^2, \ \  m_a^2 \equiv 2g m_{a}^{*2}. 
 \ee
 	Non zero $m_{a}^2$ and hence $M_{a}^2$ are imaginary numbers.  $ M_{a}^2=m_{a}^2=0$ for  diagonal gluons. 
 The Eq.~\eqref{mr}  shows the consequence of the underlying curved 
   space on mass in this theory~\cite{111}. Mass of an off-diagonal gluon on $\mathbb{S}^4$ has become position dependent.  
   Thus, the curvature  does not alter   the infrared regime of this theory on $\mathbb{S}^4$  from that of the effective theory in quadratic gauge in the $\mathbb{R}^4$ but it affects  mass of a gluon on $\mathbb{S}^4$ to be position dependent. 
   
       	So, the  effective action in the ghost condensed phase on  $\mathbb{S}^4$ which characterizes confinement now becomes
    \be\label{5}
    S_{eff}=S_{YM} +  \int d \Omega \Big[ -\frac{1}{2\xi} (\bar{A}^a_\beta  \bar{A}_\beta^a)^2+ M_{a}^2 \bar{A}^a_\beta  \bar{A}_\beta^a\Big],
    \ee with the $ M_{a}^2 $ being position dependent as in Eq.~\eqref{mr} and 	e.g, for $SU(3)$, $M^2_3=M^2_8=0$. While for the off-diagonal gluons, $M^2_1=+im^2_1, M^2_2=-im^2_1, \  
    M^2_4=+im^2_2,  M^2_5=-im^2_2, \  M^2_6=+im^2_3, M^2_7=-im^2_3 \ (m_1^2, m_2^2, m_3^2 $ 
    are positive real$)$. This action is not BRST invariant under the transformation of a gauge field. Now,   in the dual superconductor picture, Eq.~\eqref{5} can taken to be the standard expression of the confined phase in this theory regardless of a process through which it is attained.  The particular   condensation of ghosts studied here is one of the methods which leads to confinement from deconfinement.       In the next section, we propose another mechanism to achieve   confinement  from the deconfined phase   in the given theory on $\mathbb{S}^4$ that however brings with it a little difference.
 
\section{Generalized BRST as  a mechanism for confinement  on 4-sphere }
Here we propose two different field redefinitions that can implement the same  transition between deconfined and confined phases.  The action in the deconfined phase, 
 \be\label{de} S_{eff}= \int d \Omega \left[ -\frac{1}{12} \ \bar{F}_{\alpha \beta \gamma }^a \bar{F}_{\alpha \beta \gamma }^a -\frac{1}{2\xi} (\bar{A}^a_\beta  \bar{A}_\beta^a)^2-2 \hat{\bar{c}}^a \bar{A}_\beta^a D_\beta \bar{c}^a\right] \ee
	   is invariant under the following   BRST transformations 
	  \be\label{9}
	  \delta \bar{A}^a_\beta &=& \omega   ( D_\beta- r_\alpha r_\beta D_\alpha) \bar{c}^a =  \omega \ r_\alpha i\mathcal{L}_{\alpha \beta } \bar{c}^a,\nn\\
	  \delta \bar{c}^a &=&\frac{\omega}{2} f^{abc} \bar{c}^b\bar{c}^c,\nn\\
	  \delta\hat{\bar{c}}^a &=& - \omega \frac{1}{\xi}  \bar{A}^a_\alpha  \bar{A}_\alpha^a.
	  \ee
	    Before we identify a suitable FFBRST in this case,  we outline   the procedure that generalizes the BRST.  The 
	  infinitesimal global parameter $\omega$  is made field dependent  along with  introducing a numerical parameter $\kappa \ (0\leq \kappa\leq 1) $. All the fields are then made  $\kappa$ dependent
	  so that  $\phi (x,\kappa=0)=\phi(x) $ and $\phi (x,\kappa=1)=\phi^\prime(x) $, the transformed field.
	  Symbol $\phi$ generically describes all the fields $\bar{A},\bar{c},\hat{\bar{c}}$.  The BRST transformation in Eq.~\eqref{9} is then given by
	  \be\label{infb}
	  d\phi = \delta_b[\phi(x,\kappa)]\Theta^\prime(\phi(x,\kappa))\ d\kappa
	  \ee
	  where $\Theta'$ is a finite field dependent anti-commuting parameter and 
	  $\delta_b[\phi(x,\kappa)]$ is the form of the transformation for the corresponding field as in 
	  Eq.~\eqref{9}. The
	  FFBRST is then developed by integrating Eq.~\eqref{infb} from $\kappa=0$ to $\kappa=1$ as~\cite{12}
	  \be\label{ffbrst}
	  \phi^\prime\equiv \phi(x,\kappa=1)=\phi (x,\kappa=0)+\delta_b[\phi(0)]\Theta [\phi(x)]
	  \ee
	  where $\Theta[\phi(x)] =\int_0^1 d\kappa^\prime\Theta^\prime[\phi(x,\kappa)] $.   Like usual BRST transformation, FFBRST
	  transformation leaves the effective action in Eq.~\eqref{de} invariant  but it
	   does not leave the path integral measure, ${\cal D}\phi $ invariant    since the transformation parameter is field dependent.  It produces a non-trivial Jacobian  $J$ i.e., $\mathcal{D}\phi(\kappa) \rightarrow J(\kappa)\mathcal{D}\phi(\kappa)$. 
	  This $J$  can further be cast as a local exponential  functional of fields, $ e^{iS_J}$ (where the $S_J$ is the action representing the Jacobian factor $J$) 
	  if the following condition is satisfied~\cite{12}
	  \be \label{con}
	  \int { \cal D }\phi (x,\kappa) \left [\frac{1}{J}\frac{dJ}{d\kappa}-i\frac{dS_J}{d\kappa}\right ]e^{i(S_J+\mathcal{S}_{eff})}=0.
	  \ee
	  Thus the procedure for FFBRST may be summarised in three steps as (i)  calculate the infinitesimal change in 
	  Jacobian, $\frac{1}{J}
	  \frac{dJ}{d\kappa} d\kappa $ using 
	  \begin{equation}\label{j}
	  \frac{J(\kappa)}{J(\kappa+d\kappa) }= 1-\frac{1}{J(\kappa)}\frac{dJ(\kappa)}{d\kappa}d\kappa
	  = \sum_\phi \pm \frac{\delta\phi(x,\kappa+d\kappa)}{\delta\phi(x,\kappa)}
	  \end{equation}
	  for infinitesimal BRST transformation, $+$ or $-$ sign is for Bosonic or Fermion nature of the field  $\phi$ respectively 
	  (ii)  make a suitable ansatz for $S_J$, (iii)  then  check  Eq.~\eqref{con}
	  for this ansatz and if that is consistent, finally  replace $J(\kappa)$ by $e^{iS_J}$ in the generating functional
	  \begin{equation}
	  W=\int {\cal D}\phi (x) e^{iS_{eff}(\phi)} = \int {\cal D}\phi (x,\kappa) J(\kappa)e^{iS_{eff}(\phi (x,\kappa))} .
	  \label{ww}
	  \end{equation}
	  Setting $\kappa=1$, this would then provide the new effective action $ S^\prime_{eff}=S_J+S_{eff}$.
	  \subsection{FFBRST 1:}
Let us now construct a FFBRST transformation that can transform the action in deconfined phase  to that in confined phase  inside the quadratic gauge only on 4-sphere. We  begin with introducing the batch of  new fields $\bar{W}^a $ whose BRST transformation would be decided later. They are commuting scalars.
To this end, we choose the following  finite 
field dependent parameter 
\be\label{th}
\Theta'[\phi(\kappa)]=-i  \int d \Omega \left [  \gamma_1  \bar{W}^a  \bar{A}_{\beta}^a[ ( D_{\beta} \bar{c})^a-  r_\beta r_\tau (D_\tau \bar{c})^a] + \gamma_2  \xi \lambda^a     \hat{\bar{c}}^a  \right].
\ee
The
 $\lambda^a$ are  constants to be set later, $ \gamma_1, \gamma_2$  are numbers related to FFBRST and $\xi$ is a  parameter of the quadratic gauge.  Sum over the group index $a$ is understood.  Square of the field dependent parameter, $\Theta'^2=0$. Although it seems that the parameter depends on coordinates explicitly  but it does not as $ \bar{A}_{\beta}^a  r_\beta=0$ which gets rid of the term $ r_\beta r_\tau (D_\tau \bar{c})^a$. The addition of this term is useful as we shall just see   since on sphere $\delta_b [ ( D_{\beta} \bar{c})^a-  r_\beta r_\tau (D_\tau \bar{c})^a]=0 $ and not $\delta ( D_{\beta} \bar{c})^a$. Thus, ansatz of the parameter is still in line with the expectation that it should not depend on coordinates.
There are a few qualities of this FFBRST parameter not found in the usual ones. It consists of a covariant derivative  and a new field which exists neither before nor after FFBRST  operation.  The usual FFBRST parameters are of the form $(anti \ ghost)(gauge_1  + gauge_2)$ and connect two gauge conditions in flat spacetime~\cite{13,14,15,16,17,18} whereas  parameter in Eq. \eqref{th} has entirely different structure. 

  The    change in the Jacobian $\frac{1}{J}\frac{dJ}{d\kappa} $ due to  FFBRST  parameter in Eq.~\eqref{th} as per Eq.~\eqref{j}  is the following 
\be \label{j11}
\frac{1}{J}\frac{dJ}{d\kappa} &=& - \Big( \frac{\delta \Theta'}{\delta \bar{W}^d} \delta_b \bar{W}^d +\frac{\delta \Theta'}{\delta \bar{A}_\beta^d} [(D_\beta \bar{c})^d- r_\beta r_\tau (D_\tau \bar{c})^d]  
- \frac{\delta (\Theta' gf^{def}\bar{c}^e\bar{c}^f)}{2\delta \bar{c}^d}- \frac{\delta \Theta'}{\delta \partial_\beta  \bar{c}^d}\frac{g}{2}\partial_\beta (f^{def}\bar{c}^e\bar{c}^f) 
\nn\\&+&\frac{\delta \Theta'}{\delta \hat{\bar{c}}^d} \frac{1}{\xi}  \bar{A}^d_{\beta} \bar{A}^{\beta d} \Big),\ \ \text{$d$ is summed over,  $\delta_b$ denotes BRST variation}\nn\\
&=&i\int d \Omega \Big( \gamma_1 \delta_b(\bar{W}^a)  \bar{A}_{\beta}^a[ ( D_{\beta} \bar{c})^a-  r_\beta r_\tau (D_\tau \bar{c})^a] +  \gamma_1\bar{W}^a  \delta_b(\bar{A}_{\beta}^a)[ ( D_{\beta} \bar{c})^a-  r_\beta r_\tau (D_\tau \bar{c})^a] \nn\\& - &  \gamma_1 \bar{W}^a  \bar{A}_{\beta}^a \delta_b[ ( D_{\beta} \bar{c})^a-  r_\beta r_\tau (D_\tau \bar{c})^a] +  \gamma_2    \lambda^a  \bar{A}_{\beta}^a  \bar{A}_{\beta}^a   \Big) \nn\\
&=&  i\int d \Omega \Big(\gamma_1 \delta_b(\bar{W}^a)  \bar{A}_{\beta}^a[ ( D_{\beta} \bar{c})^a-  r_\beta r_\tau (D_\tau \bar{c})^a] +  \gamma_1\bar{W}^a [ ( D_{\beta} \bar{c})^a-  r_\beta r_\tau (D_\tau \bar{c})^a]^2 \nn\\& - &  \gamma_1 \bar{W}^a  \bar{A}_{\beta}^a \delta_b[ ( D_{\beta} \bar{c})^a-  r_\beta r_\tau (D_\tau \bar{c})^a] +  \gamma_2    \lambda^a  \bar{A}_{\beta}^a  \bar{A}_{\beta}^a  \Big)
.  \ee

Now we use following identities in above expression 
\be
&& r_\beta  \bar{A}_\beta^d = 0 \ \text{in the first term,}\nn\\
&&\delta_b [ D_\beta \bar{c})^d -  r_\beta r_\tau (D_\tau \bar{c})^d)]=0  \   \text{in the third term due to nipotency}, \nn\\
&&  ( D_\beta \bar{c})^d  (D_\beta \bar{c})^d =r_\beta ( D_{\beta} \bar{c})^d r_\tau (D_\tau \bar{c})^d  =0
\ \text{in the second term}.\nn\ee 
 
Using these   identities in  Eq.~\eqref{j11}, we get the following simplification
\be 
\frac{1}{J}\frac{dJ}{d\kappa} &=& i \int  d \Omega \Big( \gamma_1 \delta_b\bar{W}^a  \bar{A}^{\beta a} ( D_{\beta} \bar{c})^a+\gamma_2 \lambda^a \bar{A}^a_{\beta} \bar{A}^{\beta a}\nn\Big).
\ee	
The clue for determining
 $\delta_b \bar{W}^d  $ will now be visible.

The $\frac{1}{J}\frac{dJ}{d\kappa} $ does not contain terms that have  $\Theta'$
as multiplicative
factor, therefore the dependence on $\kappa$  of $S_J (\kappa)$ is multiplicative~\cite{12}.
This suggests that  fields in the ansatz for the $S_J$ are $\kappa$ independent. Therefore,    the
ansatz for the $S_J$ representing Jacobian would be as given below  
\be 
S_J[\phi(\kappa), \kappa] &=& \int d \Omega\Big(\alpha_1(\kappa)\ \delta_b\bar{W}^a  \bar{A}^{\beta a} ( D_{\beta} \bar{c})^a+\alpha_2(\kappa)\ \lambda^a \bar{A}^a_{\beta} \bar{A}^{\beta a}\Big)
\ee	where $\alpha_1(\kappa), \alpha_2(\kappa) $ are arbitrary functions of $\kappa$ with initial condition $\alpha_i(\kappa=0)=0$ and   fields explicitly do not depend upon $\kappa$.
Condition in Eq.~\eqref{con} in this case becomes
\be
\int {\cal D}\phi[x,\kappa]\int d\Omega \Big( [\dot{\alpha_1}(\kappa)-\gamma_1]\ \delta_b\bar{W}^a  \bar{A}^{\beta a} ( D_{\beta} \bar{c})^a+ [\dot{\alpha_2}(\kappa)-\gamma_2] \lambda^a \bar{A}^a_{\beta} \bar{A}^{\beta a}\nn \nn \Big)
e^{i(S_{eff}+S_J)}=0,
\ee
which gives the following relation among parameters
\begin{eqnarray}\label{tr}
\begin{split}
{\alpha_1}&=  \gamma_1 .\kappa\\
{\alpha_2}&= \gamma_2.\kappa.\\
\end{split}
\end{eqnarray}
We choose arbitrary parameters   $\gamma_1 =2, \gamma_2 =1$  in Eq.~\eqref{tr}.
Thus, the additional Jacobian contribution at $\kappa=1$ is
\be\label{Sj}
S_J &=& \int  d \Omega\Big( 2\    \delta_b\bar{W}^a  \bar{A}^{\beta a} ( D_{\beta} \bar{c})^a+ \lambda^a \bar{A}^a_{\beta} \bar{A}^{\beta a}\nn \Big).
\ee
Adding this Jacobian action, $S_J$ to the $\mathcal{S}_{eff}$ in Eq.~\eqref{de} we get at $\kappa=1$ the following
\be
\mathcal{S}_{eff}+S_J 
&=	&\int  d \Omega\Big[\frac{-1}{12} \bar{F}_{\alpha \beta \gamma }^a \bar{F}_{\alpha \beta \gamma }^a -\frac{1}{2\xi} (\bar{A}^a_\beta  \bar{A}_\beta^a)^2 + 2 (\delta_b\bar{W}^a - \hat{\bar{c}}^a) \bar{A}^{\beta a} ( D_{\beta} \bar{c})^a+ \lambda^a \bar{A}^a_{\beta} \bar{A}^{\beta a}\Big]
.	\ee	
There are  two  important points to be noted here, (i) the new set of fields $\bar{W}^a$ does not appear in this final action, (ii) if we choose the following BRTS transformation of $\bar{W}^a$,  the third term 	vanishes, 
\be \delta_b\bar{W}^a = \hat{\bar{c}}^a.\ee
We can freely choose this transformation as  $\bar{W}^a$s appear neither in deconfined nor in confined phase. Therefore, using this transformation, we get 
\be\label{ej}
\mathcal{S}_{eff}+S_J 
&=&  \int  d \Omega\Big[-\frac{1}{12} \bar{F}_{\alpha \beta \gamma }^a \bar{F}_{\alpha \beta \gamma }^a -\frac{1}{2\xi} (\bar{A}^a_\beta  \bar{A}_\beta^a)^2  + \lambda^a \bar{A}^a_{\beta} \bar{A}^{\beta a}\Big].
\ee	
It is clear from Eq.~\eqref{ej} that $\lambda^a$s are mass squared of gluons on $\mathbb{S}^4$  i.e.,  $\lambda^a = M_a^2$.  We set    $\lambda^a = M_a^2 $  as elaborated below Eq.~\eqref{5}. The difference from the ghost condensation mechanism is that here $\lambda^a = M_a^2 $ are not position dependent in this mechanism since the  parameter $\Theta'$ is  not  coordinate dependent as mentioned earlier.  The $ M_a^2 $ are imaginary numbers in the pattern  given below Eq.~\eqref{5}.   We have therefore attained  precisely the same   confined phase action on 4-sphere as  in Eq.~\eqref{5} through a method of field redefinition with the unique parameter in Eq.~\eqref{th} and Jacobian contribution  in Eq.~\eqref{Sj}.

	  \subsection{FFBRST 2:}
	  Confinement can be achieved through one more distinct FFBRST transformation which we now explain. FFBRST parameter in this case is relatively simple and as follows
\be\label{th1}
\Theta'[\phi(\kappa)]=-i  \int d \Omega \left [  \hat{\bar{c}}^a ( \gamma_1  \bar{A}_{\alpha}^a    \bar{A}_{\alpha}^a   + \gamma_2  \xi \lambda^a    ) \right].
\ee
The
 $\lambda^a$ are  constants to be chosen later, $ \gamma_1, \gamma_2$  are numbers related to FFBRST and, $\xi$ is a gauge parameter. Moreover,  $\Theta'^2=0$.  Sum over the group index $a$ is understood. 

The    change in the Jacobian $\frac{1}{J}\frac{dJ}{d\kappa} $ owing to the FFBRST  parameter in Eq.~\eqref{th1} as per Eq.~\eqref{j}  is the following 
\be
\frac{1}{J}\frac{dJ}{d\kappa} &=& - \Big(  \frac{\delta \Theta'}{\delta \bar{A}_\beta^d} [( D_{\beta} \bar{c})^d-  r_\beta r_\tau ( D_{\tau} \bar{c})^d] 
+\frac{\delta \Theta'}{\delta \hat{\bar{c}}^d} \frac{1}{\xi}  \bar{A}^d_{\beta} \bar{A}^{\beta d} \Big),\ \ \text{$d$ is summed over}\nn\\
&=&i\int d \Omega \Big( \frac{\gamma_1}{\xi} (\bar{A}^a_\beta  \bar{A}_\beta^a)^2+2\gamma_1 \hat{\bar{c}}^a \bar{A}^{\beta a} [( D_{\beta} \bar{c})^a-  r_\beta r_\tau ( D_{\tau} \bar{c})^a]  +\gamma_2 \lambda^a \bar{A}^a_{\beta} \bar{A}^{\beta a} \Big) \nn\\
&=& i\int d \Omega \Big( \frac{\gamma_1}{\xi} (\bar{A}^a_\beta  \bar{A}_\beta^a)^2+2\gamma_1 \hat{\bar{c}}^a \bar{A}^{\beta a} ( D_{\beta} \bar{c})^a  +\gamma_2 \lambda^a \bar{A}^a_{\beta} \bar{A}^{\beta a} \Big) \ \ \text{since $r_\beta \bar{A}^{\beta d} = 0$}.  \ee

 \noindent
Now the rest of the procedure is the same as described in FFBRST 1.  We take $\gamma_1= \gamma_2=1 $. Condition in Eq.~\eqref{con} leads to  the following      Jacobian  contribution in terms of the action  $S_J$ at $\kappa=1$
\be\label{Sj1} 
S_J[\phi(\kappa), \kappa] &=& \int d \Omega\Big(\frac{1}{\xi} (\bar{A}^a_\beta  \bar{A}_\beta^a)^2 + 2\hat{\bar{c}}^a \bar{A}^{\beta a} ( D_{\beta} \bar{c})^a + \lambda^a \bar{A}^a_{\beta} \bar{A}^{\beta a}\Big)
\ee 
 Adding this Jacobian contribution, $S_J$ to the $\mathcal{S}_{eff}$ in Eq.~\eqref{de}  at $\kappa=1$ we see that ghost term  cancels and we get  the following
\be
\mathcal{S}_{eff}+S_J 
&=	&\int  d \Omega\Big[-\frac{1}{12} \bar{F}_{\alpha \beta \gamma }^a \bar{F}_{\alpha \beta \gamma }^a +\frac{1}{2\xi} (\bar{A}^a_\beta  \bar{A}_\beta^a)^2  + \lambda^a \bar{A}^a_{\beta} \bar{A}^{\beta a}\Big]
.	\ee	
It is almost the same action of confined phase in Eq. (\ref{5}) except for the sign of the gauge fixing term. To fix it, we can further apply a second
FFBRST transformation such that $\xi \rightarrow \xi' $
 in the same gauge   \cite{12}.
  As done earlier,     $\lambda^a = M_a^2 $ is set as elaborated below Eq.~\eqref{5}.  Please note again that masses in this process are not position dependent. We have thus got  confinement on $\mathbb{S}^4$   using a different  field transformation  with the unique parameter in Eq.~\eqref{th1} and Jacobian contribution as in Eq.~\eqref{Sj1}.

\section{conclusion}
We proposed a novel process in two unique  field redefinitions  to achieve confinement from the deconfined phase on 4-sphere within the quadratic gauge besides the  ghost condensation  mechanism. There is also a difference in the final outcome of  the ghost condensation and that of  field transformation, which is that   in the later case mass of a gluon on 4-sphere is not position dependent since the field dependent parameter is    coordinate independent. The FFBRST parameters act as order parameters of the transition. In FFBRST 1, the  method is required to be extended to include a coavariant derivative as one of the grassmann variables and a new field that does not exist in any of phases. FFBRST 2 is relatively simple but has a disadvantage that it uses one more subsequent FFBRST that changes  $\xi \rightarrow \xi' $  in the same gauge to get the exact action in the confined phase.

The passages illustrated here are from phase having BRST invariance to   phase without BRST invariance, which stand out among usual cases.  
The reverse  phase  transition from the confined  to deconfined phase is not feasible as the former phase is not BRST invariant and FFBRST  operates  consistently upon BRST invariant actions only. So, both the FFBRSTs only implement  deconfinement to confinement  phase transition on 4-sphere   just like  the mechanism of ghost condensation. The field dependent parameters in Eqs.~\eqref{th}, \eqref{th1} act as  order parameters of the transition.

\end{document}